\documentclass[pre,nofootinbib,amsmath,amssymb,twocolumn,showpacs]{revtex4}
\usepackage[T1]{fontenc} \usepackage[latin1]{inputenc}
\usepackage{amsmath} \usepackage{graphicx} \usepackage{amssymb}

\makeatletter

\usepackage{subfigure} \usepackage{amsfonts}

\makeatother
\begin{document}

\title{Response of Single Polymers to Localized Step Strains}

\author{Debabrata Panja}

\affiliation{Institute for Theoretical Physics, Universiteit van
Amsterdam, Valckenierstraat 65, 1018 XE Amsterdam, The Netherlands}

\begin{abstract} In this paper, the response of single
three-dimensional phantom and self-avoiding polymers to localized step
strains are studied for two cases in the absence of hydrodynamic
interactions: (i) polymers tethered at one end with the strain created
at the point of tether, and (ii) free polymers with the strain created
in the middle of the polymer. The polymers are assumed to be in their
equilibrium state before the step strain is created. It is shown that
the strain relaxes as a power-law in time $t$ as $t^{-\eta}$. While
the strain relaxes as $1/t$ for the phantom polymer in both cases; the
self-avoiding polymer relaxes its strain differently in case (i) than
in case (ii): as $t^{-(1+\nu)/(1+2\nu)}$ and as $t^{-2/(1+2\nu)}$
respectively. Here $\nu$ is the Flory exponent for the polymer, with
value $\approx0.588$ in three dimensions. Using the mode expansion
method, exact derivations are provided for the $1/t$ strain relaxation
behavior for the phantom polymer. However, since the mode expansion
method for self-avoiding polymers is nonlinear, similar theoretical
derivations for the self-avoiding polymer proves difficult to
provide. Only simulation data are therefore presented in support of
the $t^{-(1+\nu)/(1+2\nu)}$ and the $t^{-2/(1+2\nu)}$ behavior. The
relevance of these exponents for the anomalous dynamics of polymers is
also discussed.
\end{abstract} 

\maketitle

\section{Introduction\label{sec1}}

If a polymer is subjected to local step strain, i.e., a small part of
a polymer is made to undergo a relatively fast conformational change,
during subsequent evolution the polymer will readjust itself in an
attempt to relieve its strain. The local conformational change will
alter the polymer's local chain tension; and the new chain tension
will be unable to maintain the polymer in equilibrium. In response to
that, monomers will be pulled from (or pushed away to) the adjacent
part of the polymer, thereby spreading the effect of the local
strain. In time, the effect of the  local strain will spread through
the entire polymer along its backbone, before equilibrium conditions
can be finally restored.

Studies on strain relaxation in collective polymeric systems are
abundant in traditional polymer physics, such as for
(dilute/semi-dilute) polymer solutions and for polymer melts
\cite{doi}. From this perspective, how a single polymer relieves its
local step strain may seem to be a purely theoretically motivated
problem. However, experimentalists' ability to manipulate polymeric
systems at single polymer level --- specially in the context of
biological polymers, or biopolymers --- have rapidly grown in the last
few years; e.g., DNA separation in nanochannels \cite{dnachannel},
dynamics of RNA polymerase \cite{rnap}, biopolymer translocation
\cite{expt1,expt2,expt3,expt4,expt5,expt6}, packaging and ejection of
bacteriophage DNA during infection \cite{pack,eject}, surface
desorption of polymers using a pulling force \cite{desorb}. Such
single polymer experiments have been continuously challenging polymer
theorists; one can almost claim that polymer physics at a single
polymer level is being reborn through these recent
developments. Indeed, our motivation to study the response of single
polymers to localized step strains, stem from the fact that there are
systems whose dynamics are determined by the polymers' local strain
relaxation mechanism. Take for example polymer translocation, where
the polymer passes through a narrow pore in a membrane
\cite{expt1,expt2,expt3,expt4,expt5,expt6}. A translocating polymer is
composed of two polymer strands (labeled A and B respectively), one on
each side of the membrane. The only way the two strands interact with
each other is through the pore: as the monomers translocate, they
leave one strand to join the other. Monomers leaving strand A locally
increases the chain tension of strand A at the pore, and as they join
strand B across the membrane, they reduce the chain tension of strand
B, also locally at the pore. How the segments relieve these local
strains determines the dynamics of translocation
\cite{panja1,panja2,panja3,panja4}. Similarly, in the case of polymer
adsorption on a rigid surface, when a monomer gets adsorbed, it
creates a local (at the adsorbing surface) step strain in the polymer,
and the adsorption kinetics is governed by how the polymer relieves
this strain \cite{tolya}.

The fact that local step strain relaxations of a polymer is governed
by a power-law in time can be argued on general theoretical
grounds. Let us consider the application of the step strain of
magnitude $\epsilon_0$ at a given location (say the $n^*$-th monomer)
of a polymer of length $N$ at $t=0$. This strain will excite {\it
all\/} fluctuation modes of the polymer. The amplitude $a_q$ of the
$q$-th mode $\psi_q$ can be obtained from the equation
$\epsilon_0=\sum_q a^{(0)}_q\psi_q$, $q=1,2,\ldots N$. Typically, in
polymer physics, the $q$-th fluctuation mode of a polymer has an
associated relaxation time $\tau_q\sim(N/q)^\beta$ for some $\beta$,
where $\tau_N\sim N^\beta$ is the longest relaxation time of the
polymer, corresponding to the slowest mode $q=1$ of the polymer
($\beta=1+2\nu$ for a Rouse polymer, and $\beta=3\nu$ for a Zimm
polymer). The subsequent evolution of this strain will then be given
by $\epsilon(t)=\sum_q a^{(0)}_q\psi_q\exp(-t/\tau_q)$. The local
contribution of these summed over large number of exponentials at
$n^*$ will yield a power-law, implying that $\epsilon_{n^*}(t)\sim
t^{-\eta}$ for some $\eta$, multiplied by the overall terminal
exponential decay $\sim\exp(-t/\tau_N)$. Such power-laws are often
referred to as ``memory effects''. The quantity $\eta$, the exponent
for the power-law, characterizes the response of single polymers to
local step strains. For the two physical systems discussed above,
namely polymer translocation and adsorption of polymers on rigid
surfaces, it is the exponent $\eta$ that dictates the dynamics
\cite{panja1,panja2,panja3,panja4,tolya}.

The purpose of this paper is to report the exponent $\eta$ for phantom
and self-avoiding polymers in three dimensions in the absence of
hydrodynamic interactions. The specific way we create the local strain
in the polymers is as follows. At a given location (say the $n^*$-th
monomer) of an equilibrated polymer of length $N$, we inject $p\,(\ll
N)$ crumpled monomers at $t=0$, bringing its length to
$N+p$. Following the monomer injection at $t=0$, apart from the newly
injected monomers, the polymer  follows random walk (or self-avoiding
walk) statistics, i.e., the  strain in the polymer is localized at
monomer number $n^*$. In the subsequent evolution of the polymer, we
then keep track of how these $p$ crumpled monomers unfold themselves,
which yields us the exponent $\eta$. Note that the specific way we
choose to create the local strain in the polymers is indeed motivated
by the actual microscopic dynamics of polymer translocation or polymer
adsorption on a rigid surface: as remarked above, for polymer
translocation it is the addition or disappearance of monomers to the
polymer segments on either side of the membrane that creates the local
strain (and similarly for the case of polymer adsorption on a rigid
surface).

We calculate $\eta$ for two different cases each for three-dimensional
phantom and self-avoiding polymers: (i) polymers tethered at one end
with the strain created at the point of tether, and (ii) free polymers
with the strain created in the middle of the polymer. We derive that
$\eta=1$ in both cases; however, for the self-avoiding polymer we show
that $\eta=(1+\nu)/(1+2\nu)$ for case (i), and $\eta=2/(1+2\nu)$ for
case (ii). Here $\nu$ is the Flory exponent for the polymer, with
value $\approx0.588$ in three dimensions. We provide exact derivations
for the $1/t$ strain relaxation behavior for the phantom polymer using
the mode expansion method. The mode expansion method for a
self-avoiding polymer is nonlinear, and hence similar theoretical
derivation for $\eta$ for the self-avoiding polymer proves difficult
to provide. Only high-precision simulation data are therefore
presented in support of the $t^{-(1+\nu)/(1+2\nu)}$ and the
$t^{-2/(1+2\nu)}$ step strain-relaxation behaviors of the
self-avoiding polymer.

Although the problem of local step strain relaxation behavior in the
polymers is motivated in this paper in view of polymer translocation
and polymer adsorption, note that both physical processes correspond
to the case (i) while the tether point lies on a rigid surface. The
presence of the surface, in principle, can influence the strain
relaxation mechanism, and alter the value of $\eta$ from its value in
the absence of the surface. However, since in
Refs. \cite{panja1,panja2,panja3} it was shown --- using a model that
allowed direct observation of the local strain relaxation --- that
$\eta=(1+\nu)/(1+2\nu)$ for a self-avoiding polymer for the case of
(i) in the presence of a rigid surface as well, the result of this
paper therefore implies that the local strain release mechanism for
self-avoiding tethered polymers is unaffected by the presence of a
surface at the tether point. Note that recently, albeit indirectly, a
different polymer model has confirmed that $\eta=(1+\nu)/(1+2\nu)$ for
a self-avoiding polymer for the case of (i) in the presence of a rigid
surface \cite{luo,fyta}, in support of
Refs. \cite{panja1,panja2,panja3}. 

This paper is organized as follows. In Sec. \ref{sec2a} we use the
mode expansion technique for a phantom polymer for the case of (i) and
derive that $\eta=1$. In Sec. \ref{sec2b}, we then consider case (ii)
for a phantom polymer to again derive that $\eta=1$. In
Sec. \ref{sec3} we report the corresponding results for self-avoiding
polymers, and discuss the reasons why the self-avoiding behaves
differently in case (i) than in case (ii). The paper is then concluded
in Sec. \ref{sec4} with a  discussion on the relevance of these
exponents for the anomalous dynamics of polymers.

\section{Response of phantom polymers to local step
strain\label{sec2}}

With $\vec{r}(n,t)$ as the physical location of the $n$-th monomer of
the polymer at time $t$, we start with the Rouse equation for a
phantom polymer and add thermal noise $\vec{f}(n,t)$ to it:
\begin{eqnarray} \frac{\partial \vec{r}}{\partial t}=\frac{\partial^2
\vec{r}}{\partial n^2}+\vec{f}(n,t)\,.
\label{e1}
\end{eqnarray} In Eq. (\ref{e1}) the thermal noise $\vec{f}(n,t)$
satisfies the property that $\langle\vec{f}(n,t)\rangle=0$ and
$\langle{f}_\alpha(n,t){f}_\beta(n',t')\rangle=2\delta_{\alpha\beta}\delta(n-n')\delta(t-t')$;
$\alpha,\beta=x,y,z$. For case (i), the polymer with its zeroth
monomer tethered at the origin we define the $q$-th mode for a polymer
of length $(N+p)$, tethered to a fixed point at the origin as
\cite{doi} 
\begin{eqnarray} \vec{X}_q(t)=\frac{1}{N+p}\int_0^{N+p} dn\, \sin(k_q
n)\,\vec{r}(n,t),
\label{e2}
\end{eqnarray} with $\displaystyle{k_q=\frac{\pi(2q+1)}{2(N+p)}}$, and
$q=1,2,3,\ldots$, and similarly $\vec{f}_q$, the $q$-th mode for the
thermal noise. The sine-expansion in Eq. (\ref{e2}) satisfies the
boundary condition that $\vec{r}(0,t)=0$ $\forall t$, and also that at
the free end $\displaystyle{\frac{\partial \vec{r}(n,t)}{\partial
n}}\bigg|_N=0$. For case (ii) we define the $q$-th mode for a polymer
of length $(N+p)$, moving freely in space as \cite{doi}
\begin{eqnarray} \vec{X}_q(t)=\frac{1}{N+p}\int_0^{N+p} dn\, \cos(k_q
n)\,\vec{r}(n,t),
\label{e3}
\end{eqnarray} with $\displaystyle{k_q=\frac{\pi q}{(N+p)}}$, and
$q=0,1,2,3,\ldots$, and similarly $\vec{f}_q$, the $q$-th mode for the
thermal noise. In this case the cosine-expansion satisfies the
boundary condition that that at the free ends of the polymer
$\displaystyle{\frac{\partial \vec{r}(n,t)}{\partial
n}}\bigg|_0=\displaystyle{\frac{\partial \vec{r}(n,t)}{\partial
n}}\bigg|_N=0$.

In terms of the transforms (\ref{e2}) and (\ref{e3}) the Rouse
equation (\ref{e1}) reduces to the Langevin form
\begin{eqnarray} \frac{\partial \vec{X}_q}{\partial
t}=-k^2_q\vec{X}_q+\vec{f}_q\,,
\label{e4}
\end{eqnarray} where $\vec{f}_q$ is defined similar to Eq. (\ref{e2})
[resp. Eq. (\ref{e3})]. This reduction to the Langevin form also
yields
\begin{eqnarray} \langle f_{p\alpha}(t)\rangle=0;\,\langle
f_{p\alpha}(t)f_{q\beta}(t')\rangle\!=\!\frac{1}{N+p}\,\delta_{pq}\,\delta_{\alpha\beta}\,\delta(t-t')\,.
\label{e5}
\end{eqnarray} In terms of $\vec{X}_q(t)$ the monomer locations in
physical space are then given by
\begin{eqnarray} \vec{r}(n,t)=2\sum_q \sin(k_q
n)\,\vec{X}_q(t)\quad\mbox{and}\nonumber\\&&\hspace{-5cm}\vec{r}(n,t)=2\sum_q
\cos(k_q n)\,\vec{X}_q(t).
\label{e6}
\end{eqnarray} for the end-tethered and free polymers respectively.

\subsection{Local strain relaxation for case (i): end-tethered phantom
polymers\label{sec2a}}

As we crumple the extra $p\,(\ll N)$ monomers at the tether point to
an equilibrated polymer of length $N$ at time $t=0$, the length of the
polymer instantaneously becomes $N+p$. The ensuing time-evolution of
the polymer is then described by
\begin{eqnarray}
\vec{X}_q(t)=e^{-k^2_q\,t}\,\vec{X}_q(0)\,+\,\int_0^tdt'\,e^{-k^2_q\,(t-t')}\,\vec{f_q}(t')\,,
\label{e7}
\end{eqnarray} i.e.,
\begin{widetext}
\begin{eqnarray} \hspace{-3mm}\vec{r}(n,t)=2\sum_q \sin(k_q
n)\left[e^{-k^2_q\,t}\,\vec{X}_q(0)\,+\,\int_0^tdt'\,e^{-k^2_q\,(t-t')}\,\vec{f_q}(t')\right].
\label{e8}
\end{eqnarray} After the injection of $p$ monomers at $t=0$, to follow
the deviation from random-walk statistics along the polymer's backbone
at a given location of the polymer, say at monomer number $n_0$, we
consider another nearby monomer $n_1$, define $n=|n_1-n_0|$ and
$r^2(n,t)=[\vec r(n_1,t)-\vec r(n_0,t)]\cdot[\vec r(n_1,t)-\vec
r(n_0,t)]$
\begin{eqnarray} r^2(n,t)=4\sum_{q,q'} \left\{\underbrace{[\sin(k_q
n_1)-\sin(k_q
n_0)]}_{A_q(n_1,n_0)}\left[e^{-k^2_q\,t}\,\vec{X}_q(0)\,+\,\int_0^tdt'\,e^{-k^2_q\,(t-t')}\,\vec{f_q}(t')\right]\right\}\nonumber\\
&&\hspace{-9cm}\cdot\left\{\underbrace{[\sin(k_{q'} n_1)-\sin(k_{q'}
n_0)]}_{A_{q'}(n_1,n_0)}\left[e^{-k^2_{q'}\,t}\,\vec{X}_{q'}(0)\,+\,\int_0^tdt''\,e^{-k^2_{q'}\,(t-t'')}\,\vec{f_{q'}}(t'')\right]\right\}.
\label{e9}
\end{eqnarray} With the aid of Eq. (\ref{e5}), for a {\it given\/}
polymer realization at $t=0$, the average over the evolution histories
(i.e., noise realizations) for $t>0$, denoted by the angular brackets
$\langle.\rangle$, for this polymer yields
\begin{eqnarray} \langle r^2(n,t)\rangle=4\sum_{q,q'}
\left\{A_q(n_1,n_0)A_{q'}(n_1,n_0)\,e^{-(k^2_q+k^2_{q'})t}\,[\vec{X}_q(0)\cdot\vec{X}_{q'}(0)]\right\}+\frac{6}{(N+p)}\sum_q\frac{A^2_q(n_1,n_0)}{k^2_q}\,\left[1-e^{-2k^2_qt}\right].
\label{e10}
\end{eqnarray} At $t\rightarrow\infty$, the $t$-dependent terms drop
out, leaving us with
\begin{eqnarray} \langle
r^2(n,t\rightarrow\infty)\rangle=\frac{6}{(N+p)}\sum_q\frac{[\sin(k_q
n_1)-\sin(k_q n_0)]^2}{k^2_q}\,\approx\frac{6}{\pi}\int_0^\infty
dx\,\frac{[\sin(n_1x)-\sin(n_0x)]^2}{x^2}=3n\,,
\label{e11}
\end{eqnarray} which confirms that the polymer returns to equilibrium
as $t\rightarrow\infty$, as it should.

Since the strain at $t=0$ is created at the tether point, i.e., at
monomer number zero of the polymer (of length $N+p$), to quantify its
relaxation we track $\langle||r^2(n,t)||\rangle$ by choosing
$n_0=n^*=0$ and $n_1=n$, with $n\sim O(p)$. Here $||.||$ denotes a
second average over equilibrated configurations of the polymers at
$t=0$. From Eqs. (\ref{e9}) and (\ref{e5}), we can then write
\begin{eqnarray} \langle||r^2(n,t)||\rangle=3n+4\sum_{q,q'}
\left\{\sin(k_q n)\sin(k_{q'}
n)\,e^{-(k^2_q+k^2_{q'})t}\,||\vec{X}_q(0)\cdot\vec{X}_{q'}(0)||\right\}-\frac{6}{(N+p)}\sum_q\frac{\sin^2(k_q
n)}{k^2_q}\,e^{-2k^2_qt}\,.
\label{e12}
\end{eqnarray} Notice that if the polymer of length $(N+p)$ were
already at equilibrium at $t=0$ (i.e., no step-strain were created
anywhere in the polymer), then it would have remained in equilibrium
$\forall t>0$; i.e.,
$\langle||r^2(n,t)||\rangle\equiv\langle||r^2(n,t)||^{\mbox{\scriptsize(eq)}}\rangle=3n\,\,\forall
t$. In that case, Eq. (\ref{e12}) would reduce to
\begin{eqnarray} 4\sum_{q,q'} \left\{\sin(k_q n)\sin(k_{q'}
n)\,e^{-(k^2_q+k^2_{q'})t}\,||\vec{X}^{\mbox{\scriptsize(eq)}}_q(0)\cdot\vec{X}^{\mbox{\scriptsize(eq)}}_{q'}(0)||\right\}=\frac{6}{(N+p)}\sum_q\frac{\sin^2(k_q
n)}{k^2_q}\,e^{-2k^2_qt},
\label{e13}
\end{eqnarray} where $\vec{X}^{\mbox{\scriptsize(eq)}}_q(0)$ is
obtained from Eq. (\ref{e2}) for the polymer at equilibrium at
$t=0$. An explicit calculation of Eq. (\ref{e13}) has also been
provided in Appendix A [Eqs. (\ref{eA1}-\ref{eA5})].

Based on Eq. (\ref{e13}) we can now replace the last term on the
r.h.s. of Eq. (\ref{e12}) by the l.h.s. of Eq. (\ref{e13}) to write
\begin{eqnarray} \langle||r^2(n,t)||\rangle-3n=4\sum_{q,q'}\!\sin(k_q
n)\sin(k_{q'} n)\,e^{-(k^2_q+k^2_{q'})t}\,g_{q,q'}\!,
\label{e14}
\end{eqnarray} with
$g_{q,q'}=\displaystyle{\underbrace{||\vec{X}_q(0)\cdot\vec{X}_{q'}(0)||}_{g^{(1)}_{q,q'}}-\underbrace{||\vec{X}^{\mbox{\scriptsize(eq)}}_q(0)\cdot\vec{X}^{\mbox{\scriptsize(eq)}}_{q'}(0)||}_{g^{(2)}_{q,q'}}}$.
The quantity $g^{(2)}_{q,q'}$ has already been simplified in
Eq. (\ref{e13}) as
\begin{eqnarray}
g^{(2)}_{q,q'}=\frac{3}{(N+p)}\frac{1}{2k_qk_{q'}}\,\delta_{k_q,k_{q'}},
\label{e15}
\end{eqnarray} while the quantity $g_{q,q'}^{(1)}$ is explicitly
evaluated in Appendix B [Eqs. (\ref{e16}-\ref{e19})]. Having combined
these two quantities, in the limit of $p\rightarrow0$ we find that
\begin{eqnarray} g_{q,q'}\approx-\frac{3p}{(N+p)^2k_qk_{q'}},
\label{e19a}
\end{eqnarray} which, when used in conjunction with Eqs. (\ref{e12})
and (\ref{e14}), we obtain
\begin{eqnarray}
\langle||r^2(n,t)||\rangle=3n-\frac{12p}{(N+p)^2}\sum_{q,q'}\frac{\sin(k_q
n)\sin(k_{q'}
n)\,e^{-(k^2_q+k^2_{q'})t}}{k_qk_{q'}}=3n-\frac{12p}{\pi^2}\left[\int_0^\infty
dx \frac{\sin(nx)\,e^{-x^2t}}{x}\right]^2\!\!\approx
3n\!-\!\frac{3np}{\pi t}
\label{e20}
\end{eqnarray} at long times. In other words, the local strain at the
tether point relaxes as $1/t$; i.e., the local step strain relaxation
exponent $\eta=1$.

\subsection{Local strain relaxation for case (ii): free phantom
polymers\label{sec2b}}

For the local strain relaxation following the injection $p$ crumpled
monomers at $n^*=N/2$ into freely moving phantom polymer at $t=0$ we
follow the same route as in Sec. \ref{sec2a}; however, one needs to
replace the sine-expansion by cosine-expansion. While
Eqs. (\ref{e7}-\ref{e12}) are trivially reproduced with this
replacement, for the rest of the calculation we need two small
modifications. The first one of them is to choose $n_1=(N+p-n)/2$ and
$n_0=(N+p+n)/2$ such that $\langle||r^2(n,t)||\rangle$, as defined
above Eq. (\ref{e9}), can once again quantify the local strain
relaxation of the polymer. The second one is that $A_q(n_1,n_0)$ is
now defined as $A_q(n_1,n_0)=[\cos(k_q n_1)-\cos(k_q n_0)]$. These
lead us to the equivalent forms of Eqs. (\ref{e12}-\ref{e13}) as
\begin{eqnarray}
\langle||r^2(n,t)||\rangle=3n+4\sum_{q,q'}A_q(n_1,n_0)A_{q'}(n_1,n_0)
\left\{\,e^{-(k^2_q+k^2_{q'})t}\,||\vec{X}_q(0)\cdot\vec{X}_{q'}(0)||\right\}-\frac{6}{(N+p)}\sum_q\frac{A^2_q(n_1,n_0)}{k^2_q}\,e^{-2k^2_qt}\,.
\label{e21}
\end{eqnarray} and [as explicitly evaluated in
Eqs. (\ref{eA6}-\ref{eA11}) in Appendix A]
\begin{eqnarray} 4\sum_{q,q'}A_q(n_1,n_0)A_{q'}(n_1,n_0)
\left\{\,e^{-(k^2_q+k^2_{q'})t}\,||\vec{X}^{\mbox{\scriptsize(eq)}}_q(0)
\cdot\vec{X}^{\mbox{\scriptsize(eq)}}_{q'}(0)||\right\}=\frac{6}{(N+p)}\sum_q\frac{A^2_q(n_1,n_0)}{k^2_q}\,e^{-2k^2_qt}\,.
\label{e22}
\end{eqnarray} Similarly, analogous to Eq. (\ref{e14}) we have
\begin{eqnarray}
\langle||r^2(n,t)||\rangle-3n=4\sum_{q,q'}A_q(n_1,n_0)A_{q'}(n_1,n_0)\,e^{-(k^2_q+k^2_{q'})t}\,g_{q,q'},
\label{e23}
\end{eqnarray} where
$g_{q,q'}=\displaystyle{\underbrace{||\vec{X}_q(0)\cdot\vec{X}_{q'}(0)||}_{g^{(1)}_{q,q'}}-\underbrace{||\vec{X}^{\mbox{\scriptsize(eq)}}_q(0)\cdot\vec{X}^{\mbox{\scriptsize(eq)}}_{q'}(0)||}_{g^{(2)}_{q,q'}}}$,
with
\begin{eqnarray}
g^{(2)}_{q,q'}=\frac{3}{(N+p)}\frac{1}{2k_qk_{q'}}\,\delta_{k_q,k_{q'}}.
\label{e24}
\end{eqnarray} from Eq. (\ref{e22}). The explicit evaluation of
$g_{q,q'}^{(1)}$ is carried out in Appendix B
[Eqs. (\ref{e25}-\ref{e27})]. Having combined $g_{q,q'}^{(1)}$ and
$g_{q,q'}^{(2)}$, below we present the final result for $g_{q,q'}$ in
the limit of $p\rightarrow0$:
\begin{eqnarray}
g_{q,q'}\approx-\frac{3p}{(N+p)^2}\Bigg[\frac{\sin[k_q(N+p)/2]\sin[k_{q'}(N+p)/2]}{k_qk_{q'}}\Bigg],
\label{e28}
\end{eqnarray} which, when used in conjunction with Eq. (\ref{e23}),
we obtain
\begin{eqnarray}
\langle||r^2(n,t)||\rangle=3n-\frac{12p}{(N+p)^2}\sum_{q,q'}\frac{A_{q}(n_1,n_0)A_{q'}(n_1,n_0)\sin[k_q(N+p)/2]\sin[k_{q'}(N+p)/2]\,e^{-(k^2_q+k^2_{q'})t}}{k_qk_{q'}}.
\label{e29}
\end{eqnarray} Finally, with $A_{q}(n_1,n_0)=2\sin[k_q(N+p)/2]\sin[k_q
n/2]$, and $\sin[k_q(N+p)/2]=\sin[\pi q/2]$ for $q=1,2,3,\ldots$,
Eq. (\ref{e21}) reduces to
\begin{eqnarray}
\langle||r^2(n,t)||\rangle=3n-\frac{48p}{(N+p)^2}\left[\sum_{q}\frac{\sin(k_qn)\sin^2[k_q(N+p)/2]\,e^{-k^2_qt}}{k_q}\right]^2\nonumber\\&&\hspace{-9.3cm}=3n-\frac{48p}{(N+p)^2}\left[\sum_{q\,\in\,\mbox{\scriptsize
odd}}\frac{\sin(k_qn)\,e^{-k^2_qt}}{k_q}\right]^2=3n-\frac{24p}{\pi^2}\left[\int_0^\infty
dx \frac{\sin(nx)\,e^{-x^2t}}{x}\right]^2\!\!\approx 3n-\frac{6np}{\pi
t},
\label{e31}
\end{eqnarray} which, just like Eq. (\ref{e20}), approaches its
asymptotic value $3n$ as $1/t$; i.e., once again the local step strain
relaxation exponent $\eta=1$.

\end{widetext}

\section{Response of self-avoiding polymers to local step
strain\label{sec3}}

We use a Monte Carlo based lattice polymer model to study the local
step-strain relaxation for self-avoiding polymers. In this model, the
polymer consists of a sequential chain of monomers, living on a FCC
lattice. Monomers adjacent in the string are located either in the
same, or in neighboring lattice sites. Multiple occupation of lattice
sites is not permitted, except for a set of adjacent monomers. The
polymer moves through a sequence of random single-monomer hops to
neighboring lattice sites. These hops can be along the contour of the
polymer, thus explicitly providing reptation dynamics. They can also
change the contour ``sideways'', providing Rouse dynamics. The
reptation as well as the sideways moves are attempted with rate unity,
which provides us with a definition of time in this model. This model
has been used before to simulate the diffusion and exchange of
polymers in an equilibrated layer of adsorbed polymers
\cite{klein_adsorbed}, polymer translocation under a variety of
circumstances \cite{panja1,panja2,panja3,panja4,wolterink06}, and
polymer adsorption to rigid surfaces \cite{tolya}. Multiple occupation
of the same site by adjacent monomers of the polymer, in this model,
gives rise to ``stored lengths'' (see Fig. 2 of Ref.  \cite{robin} for
an illustration). Upon injection of $p$ extra monomers into the
polymer at the lattice site where the $n^*$-th monomer [$n^*=0$ and
$N/2$ for cases (i) and (ii) respectively] is located at $t=0$ the
local stored length density is immediately increased by $p$. To
measure the local strain relaxation of the polymer we therefore track
the density of stored lengths per monomer in these new $p$ monomers,
$\rho_p(t)$ as a function of time. Of course $\rho_p(t)$ would
approach some ``offset'' value $\rho_0$ as $t\rightarrow\infty$.

\begin{figure}
\begin{center}
\includegraphics[width=\linewidth]{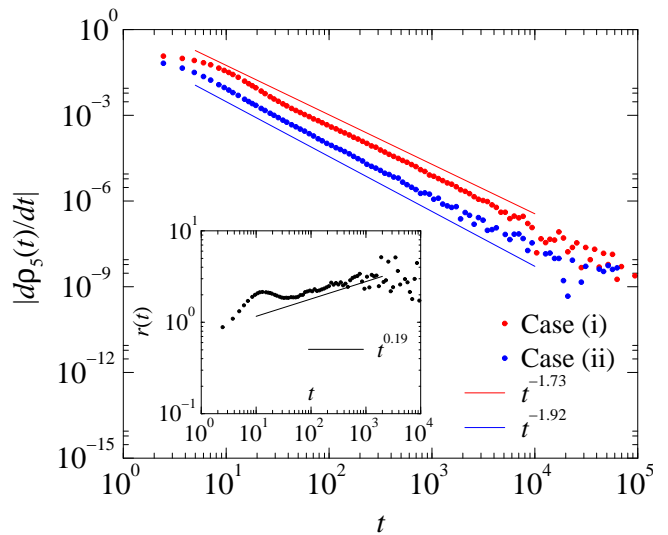}
\end{center}
\caption{(color online) Numerically differentiated data for
$|d\rho_5(t)/dt|$ for cases (i) [top set of points, in red] and (ii)
[bottom set of points, in blue], for $N=200$ and $p=5$ ($10,000,000$
realizations each), showing the respective $t^{-(1+\nu)/(1+2\nu)}$
(top straight line, in red) and $t^{-2/(1+2\nu)}$ (bottom straight
line, in blue) power-law decay for $\rho_5(t)$. Note that
$(1+\nu)/(1+2\nu)\approx0.73$ and $2/(1+2\nu)\approx0.92$. We use
numerical differentiation in order to remove the $t\rightarrow\infty$
offsets of $\rho_5(t)$. The data for case (i) is displaced upwards by
a factor 2 in the $y$-direction. Inset: Ratio $r(t)$ of the
$|d\rho_5(t)/dt|$ values for cases (i) and (ii), showing that $r(t)$
follows the power-law $t^{(1-\nu)/(1+2\nu)}$; where the value of
$(1-\nu)/(1+2\nu)$, the difference in the values of $\eta$ for cases
(i) and (ii), is $\approx0.19$. \label{fig1}}
\end{figure}

We have already argued in the introduction that the strain-relaxation
behaves as $t^{-\eta}\exp(-t/\tau_N)$. The terminal exponential decay
$\exp(t/\tau_N)$ with $\tau_N\sim N^{1+2\nu}$ is expected from the
Rouse relaxation dynamics of the entire polymer. To understand the
physics behind the exponent $\eta$, we use the well-established result
for the relaxation time $t_n$ for $n$ self-avoiding Rouse monomers
scaling as $t_n\sim n^{1+2\nu}$. On the basis of the expression of
$t_n$, we anticipate that following the injection of $p$ monomers at
$t=0$, by time $t$ the extra monomers will be well-equilibrated across
the inner part of the polymer up to $n_t\sim t^{1/(1+2\nu)}$ monomers
around $n^*$, but not significantly further. This internally
equilibrated section of $(n_t+p)$ monomers extends only to $r(n_t)\sim
n^\nu_t$, less than its equilibrated value $(n_t+p)^\nu$, because the
larger scale conformation has yet to adjust to the local strain. As a
result, internally equilibrated section of $(n_t+p)$ monomers remains
at a state of excess free energy $\delta F\sim k_{B}T[\delta
r(n_{t})/r(n_{t})]^2$. The excess $p$ monomers need to find their own
physical space by pushing the other monomers away for both cases (i)
and (ii), but for case (i) as the zeroth monomer remains tethered, we
expect them to feel a {\it force\/} of magnitude $f$ derived from the
excess free energy as $f=\partial F/\partial r(n_{t})\sim
k_{B}T\,\delta r(n_{t})/r^2(n_{t})\sim t^{-(1+\nu)/(1+2\nu)}$, which
dictates the relaxation of the step-strain; i.e.,
$\eta=(1+\nu)/(1+2\nu)$. In case (ii) however, the force derived from
the excess free energy does not yield $\eta$, as the internally
equilibrated section will simply move under the effect of the
force. Instead, in case (ii) we expect these $p$ monomers to feel a
{\it chemical potential\/} of magnitude $\mu$ derived from the excess
free energy as $\mu=\partial F/\partial n_{t}=[\partial F/\partial
r(n_{t})][\partial r(n_{t})/\partial n_{t}]\sim t^{-2/(1+2\nu)}$. The
step strain relaxation is then dictated by the chemical potential
$\mu$; i.e., $\eta=2/(1+2\nu)$.  In Fig. \ref{fig1}, by tracking
$\rho_5(t)$ for $N=195$ and $p=5$, we provide confirmation of this
physics. Note that the result for $\eta$ for case (i) is consistent
with the corresponding two dimensional case in Ref. \cite{panja3}, as
it should be.

\section{Discussion\label{sec4}}

In this paper, response of single polymers to localized step strains
is studied for two cases in the absence of hydrodynamic interactions:
(i) polymers tethered at one end with the strain created at the point
of tether, and (ii) free polymers with the strain created in the
middle of the polymer. The polymers are assumed to be in their
equilibrium state before the step strain is created. Using mode
expansion technique for Rouse equation it is shown that for phantom
polymers in both cases the strain relaxes in time as $1/t$. However,
for self-avoiding polymers for the two cases the strain relaxes as
$t^{-(1+\nu)/(1+2\nu)}$ and as $t^{-2/(1+2\nu)}$ respectively. The
strain relaxation behavior $t^{-(1+\nu)/(1+2\nu)}$ for a self-avoiding
polymer for case (i) is consistent with an earlier reported result in
two dimensions \cite{panja3}. Based on the results reported here, and
combined with those of Refs. \cite{panja1,panja4,tolya} we can
conclude that the result for case (i) is independent of the presence
of a surface at the tether point.

Although in both cases (i) and (ii) the local step strain puts the
polymer in a state of excess free energy, the difference between the
results for the self-avoiding polymers for these two cases stems from
the fact that the tether point provides a point of reference for the
polymer in case (i), but not in case (ii). As a result, for case (i)
we need to consider the force, while for case (ii) we need to consider
the chemical potential, derived from the excess free energy. For
phantom polymers however, since different parts of the polymer do not
interact with each other, there is no need for the strained monomers
to physically push away the other monomers of the polymer in order to
be able to relieve their strain, and hence for case (i), the force
derived from the excess free energy plays no role in the localized
strain relaxation for the phantom polymer. In fact, precisely because
of the same reason, we expect to see $1/t$ strain relaxation for
phantom polymers also in the presence of a surface at the tether
point. With $t^{-1}=t^{-2/(1+2\nu)}$ for phantom polymers ($\nu=0.5$),
the relevance of this paper is that one cannot trivially extend the
local strain relaxation behavior for tethered phantom polymers to
self-avoiding polymers by replacing $\nu=0.5$ by $\nu\approx0.588$ in
three dimensions.

In earlier published works  \cite{panja1,robin,panja3,panja2,panja4},
a ``voltage-current'' relationship
$\displaystyle{\phi(t)=\int_0^tdt'\mu(t-t')\dot{s}(t')}$ between
$\dot{s}(t)$, the instantaneous rate of translocation, and the
polymer's chain tension imbalance $\phi(t)$ across the pore was
established, where $\mu(t)$ is the memory effect derived from the
polymer's local strain (alternatively, the chain tension) relaxation
behavior at the pore. Here $s(t)$ is the number of the monomer located
in the pore at time $t$. Using  $\mu(t)\sim t^{-(1+\nu)/(1+2\nu)}$ for
unbiased polymer translocation \cite{panja1,robin,panja3} as in case
(i) for self-avoiding polymers in this paper, the anomalous dynamics,
characterized by $\langle\Delta s^2(t)\rangle$, where $\Delta s(t)$ is
the total number of monomers translocated through the pore in time
$t$, was then derived by using the fluctuation-dissipation theorem,
where the angular brackets denote an ensemble average. It was found
that for a translocating polymer of length $N$, $\langle \Delta
s^2(t)\rangle\sim t^{(1+\nu)/(1+2\nu)}$ up to the Rouse time
$\tau_N\sim N^{1+2\nu}$, and since no memory can survive in the
polymer beyond the Rouse time, $\langle\Delta s^2(t)\rangle\sim t$ for
$t>\tau_N$, i.e., the pore-blockade time scaling as $N^{2+\nu}$. This
result for the scaling of the pore-blockade time is in good numerical
agreement with that of Refs. \cite{dubbel,gary1}, obtained using
completely different polymer models. Furthermore, having exploited the
same ``current-voltage'' relationship between $\dot{s}(t)$ and the
chain tension difference $\phi(t)$ across the pore and that
$\mu(t)\sim t^{-(1+\nu)/(1+2\nu)}$ for field-driven translocation as
well, the exponent $N^{(1+2\nu)/(1+\nu)}$ scaling was later found for
the pore-blockade time for field-driven translocation of a polymer of
length $N$ \cite{panja4} (this result has recently been confirmed
\cite{luo} using another different polymer model). Similarly, for the
non-equilibrium dynamics of single polymer adsorption to solid
surfaces, the adsorption time for a polymer of length $N$ at weak
adsorption energies was also found to scale as $N^{(1+2\nu)/(1+\nu)}$
\cite{tolya}. These results, put together with the discussions in the
above paragraph [namely that the value of $\eta$ for case (i) is
independent of the presence of a surface at the tether point], lead us
to expect that the pore-blockade time for unbiased translocation
should scale as $N^{2+\nu}$ for self-avoiding polymers, and as $N^2$
for phantom ones, irrespective of whether translocation proceeds
through a narrow pore in a membrane or whether it proceeds through a
narrow ring (i.e., a pore without a membrane).

It is imperative to ask, based on the local strain relaxation result
for case (ii), whether it would be possible to derive an expression
for the mean-square-displacement $\langle\Delta r^2(n,t)\rangle$ of
the $n$-th monomer in physical space in time $t$, by tracking the
physical location $\vec{r}(n,t)$ for the $n$-th monomer of the polymer
at time $t$. In order to answer this question, let us reconsider the
``voltage-current'' relationship between the chain tension imbalance
across the pore and $\dot s(t)$, and note that for translocation
$s(t)$ is a scalar variable, while $\vec{r}(n,t)$ is a vector, and as
a result, deriving $\langle\Delta r^2(n,t)\rangle$ in a similar manner
is more complicated. To illustrate this difficulty, let us return to
the deterministic part of Eq. (\ref{e1}): by first expressing
$\vec{r}$ as a function of the polymer's contour $l$, and then
expressing the $l$ as a function of $n$, Eq. (\ref{e1}) reads
\begin{eqnarray} \frac{\partial \vec{r}(n,t)}{\partial
t}=\frac{\partial^2 \vec{r}}{\partial l^2} \left(\frac{\partial
l}{\partial n}\right)^2+\frac{\partial \vec{r}}{\partial
l}\,\frac{\partial^2 l}{\partial n^2}\,.
\label{e32}
\end{eqnarray} The first term on the r.h.s. of Eq. (\ref{e32}) is a
force that acts on the $n$-th monomer perpendicular to the contour of
the polymer at the location of the $n$-th monomer at time $t$, while
the second term is a force on the $n$-th monomer that acts along the
contour. Note also that the term $\displaystyle{\frac{\partial^2
l}{\partial n^2}}$ is precisely the imbalance in the chain tension
$\displaystyle{\frac{\partial l}{\partial n}}$ at the $n$-th
monomer. In the case of translocation, the fact that the motion of the
monomer perpendicular to the polymer's contour in the pore is
completely blocked means that the motion of the monomer in the pore is
determined entirely by the chain tension imbalance across the
pore. For a free polymer however, the first term on the r.h.s. of
Eq. (\ref{e32}) does contribute to the motion of the $n$-th monomer,
but what is its precise contribution to $\langle \Delta
r^2(n,t)\rangle$ is not entirely clear. Nevertheless, if we consider
the second term alone, then it does allow us to write a
voltage-current relationship (exactly the same as that of
Refs. \cite{panja1,robin,panja3,panja2,panja4}) between the chain
tension imbalance at the $n$-th monomer and the along-the-contour
velocity component of the $n$-th monomer, but this time, following the
polymer's local strain relaxation behavior for case (ii), with
$\mu(t)\sim t^{-2/(1+2\nu)}$. The application of the
fluctuation-dissipation theorem would then imply that $\langle \Delta
r^2(n,t)\rangle$ should increase as $t^{2/(1+2\nu)}$ along the
polymer's contour, i.e., in physical space $\langle \Delta
r^2(n,t)\rangle\sim t^{2\nu/(1+2\nu)}$, till the Rouse time
$\tau_N\sim N^{1+2\nu}$; this is a well-known result in polymer
physics.

\noindent {\bf Acknowledgments:} The author thanks Prof. Robin C. Ball
for helpful discussions, and Prof. G. T. Barkema for helpful
discussions as well as for the help with the numerics of
Fig. \ref{fig1}. Ample computer time from the Dutch national
supercomputer facility SARA is also gratefully acknowledged.

\begin{widetext}
\appendix
\section*{Appendix A: Derivation of
$||\vec{X}^{\mbox{\scriptsize(eq)}}_q(0)\cdot\vec{X}^{\mbox{\scriptsize(eq)}}_{q'}(0)||$
for phantom polymers}

\setcounter{equation}{0}
\renewcommand{\theequation}{A\arabic{equation}}

Here we provide a derivation of Eq. (\ref{e13}) for case (i) and an
analogous form of it for case (ii).

For case (i), by definition
\begin{eqnarray}
||\vec{X}^{\mbox{\scriptsize(eq)}}_q(0)\cdot\vec{X}^{\mbox{\scriptsize(eq)}}_{q'}(0)||=\frac{1}{(N+p)^2}\int_0^{N+p}dn\,
\sin(k_qn)\int_0^{N+p}dn'
\sin(k_{q'}n')\,||\vec{r}(n)\cdot\vec{r}(n')||^{\mbox{\scriptsize(eq)}}.
\label{eA1}
\end{eqnarray} In equilibrium the polymer satisfies random walk
statistics along its entire backbone. Hence, with $\Theta(x)$ denoting
the Heavyside function of $x$,
\begin{eqnarray}
||\vec{r}(n)\cdot\vec{r}(n')||^{\mbox{\scriptsize(eq)}}=3n\Theta(n'-n)+3n'\Theta(n-n'),
\label{eA2}
\end{eqnarray} which reduces Eq. (\ref{eA1}) to
\begin{eqnarray}
\hspace{-7mm}||\vec{X}^{\mbox{\scriptsize(eq)}}_q(0)\cdot\vec{X}^{\mbox{\scriptsize(eq)}}_{q'}(0)||\nonumber\\&&
\hspace{-3.7cm}=\frac{3}{(N+p)^2}\bigg[\int_0^{N+p}\!\!\!\!dn\,n
\sin(k_qn)\int_n^{N+p}\!\!\!\!dn'
\sin(k_{q'}n')+\int_0^{N+p}\!\!\!\!dn'\,n'
\sin(k_{q'}n')\int_{n'}^{N+p}\!\!\!\!dn \sin(k_qn)\bigg]\nonumber\\&&
\hspace{-3.7cm}=\frac{3}{(N+p)^2}\bigg[\int_0^{N+p}dn\,n
\frac{\sin(k_qn)\cos(k_{q'}n)}{k_{q'}}+\int_0^{N+p}dn'\,n'
\frac{\sin(k_{q'}n')\cos(k_qn')}{k_q}\bigg]\nonumber\\
&&\hspace{-3.7cm}=\!\frac{3}{(N\!+\!p)^2}\!\Bigg[\frac{\sin[2k_q(N+p)]-2k_q(N+p)\cos[2k_q(N+p)]}{4k_q^3}\,\delta_{k_q,k_{q'}}+(1-\delta_{k_q,k_{q'}})\times\nonumber\\&&\hspace{-2.1cm}\frac{k_q\cos[k_{q'}(N\!+\!p)]\sin[k_q(N\!+\!p)]\!-\!\cos[k_q(N\!+\!p)]\left\{(k_q^2\!-\!k_q'^2)(N\!+\!p)\cos[k_{q'}(N\!+\!p)]\!+\!k_{q'}\sin[k_{q'}(N\!+\!p)]\right\}}{k_qk_{q'}(k_q^2\!-\!k_{q'}^2)}\Bigg]\!.
\label{eA3}
\end{eqnarray} The second step of Eq. (\ref{eA3}) requires
$\cos[k_q(N+p)]=\cos[k_{q'}(N+p)]=0$, while in the last step using
$\cos[k_q(N+p)]=\cos[k_{q'}(N+p)]=0$, we first see that
$||\vec{X}^{\mbox{\scriptsize(eq)}}_q(0)\cdot\vec{X}^{\mbox{\scriptsize(eq)}}_{q'}(0)||\propto\delta_{k_q,k_{q'}}$,
and moreover, with $\sin[2k_q(N+p)]=0$ and $\cos[2k_q(N+p)]=-1$, we
obtain
\begin{eqnarray}
||\vec{X}^{\mbox{\scriptsize(eq)}}_q(0)\cdot\vec{X}^{\mbox{\scriptsize(eq)}}_{q'}(0)||=\frac{3}{(N+p)}\frac{1}{2k_q^2}\,\delta_{k_q,k_{q'}},
\label{eA4}
\end{eqnarray} i.e.,
\begin{eqnarray} 4\sum_{q,q'} \left\{\sin(k_q n)\sin(k_{q'}
n)\,e^{-(k^2_q+k^2_{q'})t}\,||\vec{X}^{\mbox{\scriptsize(eq)}}_q(0)\cdot\vec{X}^{\mbox{\scriptsize(eq)}}_{q'}(0)||\right\}=\frac{6}{(N+p)}\sum_q\frac{\sin^2(k_q
n)}{k^2_q}\,e^{-2k^2_qt}\,.
\label{eA5}
\end{eqnarray}

To derive a similar expression for
$||\vec{X}^{\mbox{\scriptsize(eq)}}_q(0)\cdot\vec{X}^{\mbox{\scriptsize(eq)}}_{q'}(0)||$
for case (ii) we express $\vec r(n,0)$, the physical location of the
$n$-th monomer at $t=0$, relative to $\vec r(0,0)$, the  physical
location of the first monomer at $t=0$ as $\vec r(n,0)=\vec
r(0,0)+\vec r\,'(n,0)$. Then
\begin{eqnarray} \vec{X}_q(0)=\frac{1}{(N+p)}\int_0^{N+p} dn\,
\cos(k_q n)\,\vec{r}(n,0)=\frac{1}{(N+p)}\int_0^{N+p} dn\, \cos(k_q
n)\,[\vec{r}(0,0)+\vec{r}\,'(n,0)],
\label{eA6}
\end{eqnarray} implying that
\begin{eqnarray}
\hspace{-5mm}||\vec{X}^{\mbox{\scriptsize(eq)}}_q(0)\cdot\vec{X}^{\mbox{\scriptsize(eq)}}_{q'}(0)||=\frac{1}{(N+p)^2}\int_0^{N+p}dn\,\cos(k_qn)\int_0^{N+p}dn'
\cos(k_{q'}n')\left[||r^2(0,0)||+||\vec{r}\,'(n)\cdot\vec{r}\,'(n')||\right]^{\mbox{\scriptsize(eq)}}\nonumber\\&&\hspace{-13.5cm}=\frac{1}{(N+p)^2}\int_0^{N+p}dn\,\cos(k_qn)\int_0^{N+p}dn'
\cos(k_{q'}n')\,||\vec{r}\,'(n)\cdot\vec{r}\,'(n')||^{\mbox{\scriptsize(eq)}}.
\label{eA7}
\end{eqnarray} To obtain the second step of Eq. (\ref{eA7})
$||r^2(0,0)||=0$ has been used by a trivial translation of origin to
obtain $\vec{r}(0,0)=0$, without affecting any part of the
calculation.

In terms of $\vec r\,'(n,0)$, we can once again use 
\begin{eqnarray}
||\vec{r}(n)\cdot\vec{r}(n')||_{p=0}=3n\Theta(n'-n)+3n'\Theta(n-n'),
\label{eA8}
\end{eqnarray} which reduces the expression for
$||\vec{X}^{\mbox{\scriptsize(eq)}}_q(0)\cdot\vec{X}^{\mbox{\scriptsize(eq)}}_{q'}(0)||$
to
\begin{eqnarray}
\hspace{-7mm}||\vec{X}^{\mbox{\scriptsize(eq)}}_q(0)\cdot
\vec{X}^{\mbox{\scriptsize(eq)}}_{q'}(0)||
\nonumber\\&&\hspace{-3.7cm}=\frac{3}{(N+p)^2}\bigg[\int_0^{N+p}dn\,n
\cos(k_qn)\int_n^{N+p}dn' \cos(k_{q'}n') +\int_0^{N+p}dn'\,n'
\cos(k_{q'}n')\int_{n'}^{N+p}dn \cos(k_qn)\bigg]\nonumber\\&&
\hspace{-3.7cm}=-\frac{3}{(N+p)^2}\bigg[\int_0^{N+p}dn\,n
\frac{\cos(k_qn)\sin(k_{q'}n)}{k_{q'}}+\int_0^{N+p}dn'\,n'
\frac{\cos(k_{q'}n')\sin(k_qn')}{k_q}\bigg]\nonumber\\
&&\hspace{-3.7cm}=\frac{3}{(N+p)^2}\Bigg[\frac{2k_q(N+p)\cos[2k_q(N+p)]-\sin[2k_q(N+p)]}{4k_q^3}\,\delta_{k_q,k_{q'}}+(1-\delta_{k_q,k_{q'}})\times\nonumber\\&&\hspace{-2.1cm}\frac{k_{q'}\cos[k_{q'}(N\!+\!p)]\sin[k_q(N\!+\!p)]-\sin[k_{q'}(N\!+\!p)]\left\{(k_q^2\!-\!k_q'^2)(N\!+\!p)\sin[k_q(N\!+\!p)]\!+\!k_q\cos[k_q(N\!+\!p)]\right\}}{k_qk_{q'}(k_q^2\!-\!k_{q'}^2)}\Bigg]\!.
\label{eA9}
\end{eqnarray} The second step of Eq. (\ref{eA9}) requires
$\sin[k_q(N+p)]=\sin[k_{q'}(N+p)]=0$, while in the last step using
$\sin[k_q(N+p)]=\sin[k_{q'}(N+p)]=0$, we first see that
$||\vec{X}^{\mbox{\scriptsize(eq)}}_q(0)\cdot\vec{X}^{\mbox{\scriptsize(eq)}}_{q'}(0)||\propto\delta_{k_q,k_{q'}}$,
and moreover, with $\sin[2k_q(N+p)]=0$ and $\cos[2k_q(N+p)]=1$, we
obtain
\begin{eqnarray}
||\vec{X}^{\mbox{\scriptsize(eq)}}_q(0)\cdot\vec{X}^{\mbox{\scriptsize(eq)}}_{q'}(0)||=\frac{3}{(N+p)}\frac{1}{2k_q^2}\,\delta_{k_q,k_{q'}},
\label{eA10}
\end{eqnarray} Equation (\ref{e10}) then yields us
\begin{eqnarray} 4\sum_{q,q'}A_q(n_1,n_0)A_{q'}(n_1,n_0)
\left\{\,e^{-(k^2_q+k^2_{q'})t}\,||\vec{X}^{\mbox{\scriptsize(eq)}}_q(0)
\cdot\vec{X}^{\mbox{\scriptsize(eq)}}_{q'}(0)||\right\}=\frac{6}{(N+p)}\sum_q\frac{A^2_q(n_1,n_0)}{k^2_q}\,e^{-2k^2_qt}\,.
\label{eA11}
\end{eqnarray}

\section*{Appendix B: Derivation of $g_{q,q'}^{(1)}$ for phantom
polymers} 

\setcounter{equation}{0}
\renewcommand{\theequation}{B\arabic{equation}}

To evaluate $g_{q,q'}^{(1)}$ for case (i) we note that
$\displaystyle{\vec{X}_q(0)=\frac{1}{N+p}\int_0^{N+p} dn\, \sin(k_q
n)\,\vec{r}(n,0)}$, and since $\vec{r}(n,0)\equiv0$ for $n\le p$ by
construction, $\displaystyle{\vec{X}_q(0)=\frac{1}{N+p}\int_0^N dn\,
\sin[k_q(n+p)]\,\vec{r}(n+p,0)}$, and hence
\begin{eqnarray}
g_{q,q'}^{(1)}=\frac{1}{(N+p)^2}\int_0^{N}dn\,\int_0^{N}dn'\,\sin[k_q(n+p)]\,\sin[k_{q'}(n'+p)]\,||\vec{r}(n+p)\cdot\vec{r}(n'+p)||.
\label{e16}
\end{eqnarray} Since the polymer was in equilibrium before the $p$
crumpled monomers were injected at the tether point, we can write 
\begin{eqnarray}
||\vec{r}(n+p)\cdot\vec{r}(n'+p)||=3n\,\Theta(n'-n)+3n'\,\Theta(n-n').
\label{e17}
\end{eqnarray} Thereafter, using Eq. (\ref{e17}), and
$\cos[k_q(N+p)]=\cos[k_{q'}(N+p)]=\sin[(k_q-k_{q'})(N+p)]=\sin[(k_q+k_{q'})(N+p)]=\sin[2k_q(N+p)]=0$
and $\cos[2k_q(N+p)]=-1$, the expression for $g_{q,q'}^{(1)}$ in
Eq. (\ref{e16}) simplifies as
\begin{eqnarray} g_{q,q'}^{(1)}=\frac{3}{(N+p)^2}\bigg[\int_0^{N}dn\,n
\sin[k_q(n+p)]\int_n^{N}dn' \sin[k_{q'}(n'+p)]+\int_0^{N}dn'\,n'
\sin[k_{q'}(n'+p)]\int_{n'}^{N}dn
\sin[k_q(n+p)]\nonumber\\&&\hspace{-17cm}=\frac{3N}{2(N+p)^2k_q^2}\delta_{k_q,k_{q'}}-\frac{3}{(N+p)^2}\frac{\sin[(k_q+k_{q'})p]}{2k_qk_{q'}(k_q+k_{q'})}-\frac{3}{(N+p)^2}\frac{\sin[(k_q-k_{q'})p]}{2k_qk_{q'}(k_q-k_{q'})}(1-\delta_{k_q,k_{q'}}).
\label{e18}
\end{eqnarray} In the limit $p\rightarrow0$ the two terms proportional
to $\delta_{k_q,k_{q'}}$ in Eq. (\ref{e18}) cancel each other, as
Eqs. (\ref{e15}) and (\ref{e18}) then leave us with
\begin{eqnarray} g_{q,q'}\approx-\frac{3p}{(N+p)^2k_qk_{q'}},
\label{e19}
\end{eqnarray}

To evaluate $g_{q,q'}^{(1)}$ for case (ii) we express $\vec r(n,0)$,
the physical location of the $n$-th monomer at $t=0$, relative to
$\vec r(0,0)$, the physical location of the first monomer at $t=0$ as
$\vec r(n,0)=\vec r(0,0)+\vec r\,'(n,0)$, to obtain
\begin{eqnarray}
\hspace{-5mm}g_{q,q'}^{(1)}=\frac{1}{(N+p)^2}\int_0^{N+p}dn\,\int_0^{N+p}dn'\,\cos(k_qn)\,\cos(k_{q'}n')\,||\vec{r}\,'(n,0)\cdot\vec{r}\,'(n',0)||\nonumber\\&&\hspace{-11.5cm}=\frac{1}{(N+p)^2}\Bigg[\int_0^{N+p}\!dn\,\int_n^{N+p}\!dn'\,\cos(k_qn)\,\cos(k_{q'}n')f(n)+\int_0^{N+p}dn'\,\int_{n'}^{N+p}dn\,\cos(k_qn)\,\cos(k_{q'}n')f(n')\Bigg]\!.
\label{e25}
\end{eqnarray} where
$f(n)=[3n\Theta(N/2-n)+3N/2\Theta(n-N/2)\Theta(N/2+p-n)+3(n-p)\Theta(n-N/2-p)]$.Thereafter,
with
$\sin[k_q(N+p)]=\sin[k_{q'}(N+p)]=\sin[(k_q-k_{q'})(N+p)]=\sin[(k_q+k_{q'})(N+p)]=0$
and $\cos[2k_q(N+p)]=1$, we find
\begin{eqnarray}
\hspace{-7mm}g_{q,q'}^{(1)}\!=\!\frac{3}{(N\!+\!p)^2}\frac{2k_qN+2\sin(k_qp)\cos[k_q(N\!+\!p)]}{4k^3_q}\delta_{k_q,k_{q'}}-\frac{3}{(N\!+\!p)^2}\times\nonumber\\&&\hspace{-10cm}\Bigg[\frac{\cos[(k_q-k_{q'})(N+p)/2]\sin[(k_q-k_{q'})p/2]}{k_qk_{q'}(k_q-k_{q'})}-\frac{\cos[(k_q+k_{q'})(N+p)/2]\sin[(k_q+k_{q'})p/2]}{k_qk_{q'}(k_q+k_{q'})}\Bigg](1-\delta_{k_q,k_{q'}}).
\label{e26}
\end{eqnarray} In the limit $p\ll N$ eq. (\ref{e26}) can be expanded
to obtain
\begin{eqnarray}
g_{q,q'}^{(1)}\approx\frac{3}{2(N\!+\!p)\,k_qk_{q'}}\delta_{k_q,k_{q'}}-\frac{3p}{(N\!+\!p)^2}\Bigg[\frac{\sin[k_q(N+p)/2]\sin[k_{q'}(N+p)/2]}{k_qk_{q'}}\Bigg].
\label{e27}
\end{eqnarray}

\end{widetext}


\begin{thebibliography}{10}

\bibitem{doi} M. Doi and S. F. Edwards, {\it The Theory of Polymer
Dynamics}, Clarendon Press, Oxford (Reprint Edition, 1999).

\bibitem{dnachannel}  J. Han and H. G. Craighead, Science {\bf 288},
1026 (2000). 

\bibitem{rnap} G. J. L. Wuite {\it et al.}, Science {\bf 287}, 2500
(2000).

\bibitem{expt1} J. Kasianowicz {\it et al.}, PNAS USA {\bf 93}, 13770
(1996).

\bibitem{expt2} S. E. Henrickson {\it et al.}, Phys. Rev. Lett. {\bf
85}, 3057 (2000).

\bibitem{expt3} A. Meller {\it et al.}, Phys. Rev. Lett. {\bf 86}, 3435
(2001).

\bibitem{expt4} M. Akeson {\it et al.}, Biophys. J. {\bf 77}, 3227
(1999).

\bibitem{expt5} A. Meller {\it et al.}, PNAS USA {\bf 97}, 1079 (2000)

\bibitem{expt6} A. Meller and D. Branton, Electrophoresis {\bf 23},
2583 (2002).

\bibitem{pack} D. E. Smith {\it et al.}, Nature {\bf 413}, 748 (2001).

\bibitem{eject} P. Grayson {\it et al.}, Proc. Natl. Sci. USA {\bf
104}, 14652 (2007).

\bibitem{desorb} C. Friedsam, M. Seitz and H. E. Gaub, J. Phys.:
Condens. Matter {\bf 16}, S2369 (2004). 

\bibitem{panja1} D. Panja, G. T. Barkema and R. C. Ball, J. Phys.:
Condens. Mattter {\bf 19}, 432202 (2007).

\bibitem{panja2} D. Panja and G. T. Barkema, Biophys. J. {\bf 94},
1630 (2008).

\bibitem{panja3} D. Panja, G. T. Barkema and R. C. Ball, J. Phys.:
Condens. Matter {\bf 20}, 075101 (2008).

\bibitem{panja4} H. Vocks {\it et al.} J. Phys.: Condens. Matter {\bf
20}, 095224 (2008).

\bibitem{tolya} D. Panja, G. T. Barkema and A. B. Kolomeisky,
arXiv:0809.0302. 

\bibitem{luo} A. Bhattacharya {\it et al.}, arXiv:0808.1868.

\bibitem{fyta} M. Fyta {\it et al.}, Phys. Rev. E {\bf 78}, 036704
  (2008).

\bibitem{klein_adsorbed} J. Klein Wolterink, G.T. Barkema and
M.A. Cohen Stuart, Macromolecules {\bf 38}, 2009 (2005).

\bibitem{wolterink06} J.K. Wolterink, G.T. Barkema and D. Panja,
Phys. Rev. Lett. {\bf 96}, 208301 (2006).

\bibitem{robin} D. Panja, G. T. Barkema and R. C. Ball,
arXiv:cond-mat/0610671.

\bibitem{dubbel} J. L. A. Dubbeldam {\it et al.}, Phys. Rev. E {\bf
76}, 010801(R) (2007).  

\bibitem{gary1} M. G. Gauthier and G. W. Slater, J. Chem. Phys. {\bf
    128}, 205103 (2008). 

\end{thebibliography}
\end{document}